\documentclass[aps,superscriptaddress, twocolumn,prl, showpacs,preprintnumbers,nofootinbib, amsmath,amssymb,
aps, superscriptaddress]{revtex4-1}

\usepackage{graphicx}
\usepackage{amsmath,amssymb}
\usepackage{amsfonts}
\usepackage{xspace} 
\usepackage[usenames]{color}
\usepackage{dcolumn}
\usepackage{bm}
\usepackage{mathrsfs}
\usepackage{multirow}
\usepackage[colorlinks=true]{hyperref}
\usepackage[all]{hypcap} 
\usepackage[utf8]{inputenc} 
\usepackage{lipsum}
\usepackage{etoolbox}
\usepackage{tikz}
\usepackage{url}
\usepackage{subfigure}
\usepackage{adjustbox}
\usepackage{caption}
\usepackage{slashed}
\usepackage{multirow}
\usepackage{array}
\usepackage{booktabs}
\usepackage{siunitx}
\usepackage{comment}
\usepackage{lineno}
\usepackage{orcidlink}
\usepackage{soul}

\def\be{\begin{equation}}
\def\ee{\end{equation}}
\def\bea{\begin{eqnarray}}
\def\eea{\end{eqnarray}}

\newcommand{\beq}{\begin{eqnarray}}
\newcommand{\eeq}{\end{eqnarray}}
\usepackage{calrsfs}
\DeclareMathAlphabet{\pazocal}{OMS}{zplm}{m}{n}

\begin{document}

\title{Black-hole evaporation from the semiclassical Einstein equations in 3+1 dimensions}

\author{Adri\'an del R\'{\i}o\,\orcidlink{0000-0002-9978-2211}}\email{adrdelri@math.uc3m.es}
 \affiliation{Universidad Carlos III de Madrid, Departamento de Matem\'aticas.\\ Avenida de la Universidad 30 (edificio Sabatini), 28911 Legan\'es (Madrid), Spain.}
 
 \author{F. Javier Mara\~n\'on-Gonz\'alez\,\orcidlink{0009-0003-8068-6910}}\email{jmarag@ific.uv.es}
\affiliation{Instituto de F\'isica Corpuscular (IFIC), CSIC-Universitat de Val\`encia and Departament de F\'isica Te\`orica. \\ Carrer del Dr Moliner 50, 46100 Burjassot (Valencia), Spain.}

\begin{abstract}

We solve the semiclassical Einstein equations for an evaporating Schwarzschild black hole in closed form, obtaining its backreacted geometry in 3+1 dimensions. This is achieved by a reformulation of the Hadamard renormalization prescription, which admits explicit stress-energy tensors for scalar quantum fields in 3+1-dimensional curved spacetimes without the standard mode-sum construction. The Unruh-like states we construct reproduce the Hawking flux at future null infinity  and the associated  ingoing negative-energy flux at the horizon, yielding  the first analytical quasi-stationary solution describing black-hole evaporation in four-dimensional semiclassical gravity. The resulting  geometry exhibits a timelike apparent horizon, providing a new analytical setting to study the causal structure of black-hole evaporation and its connection with the information-loss issue. Our results open a new route to semiclassical backreaction in four dimensions, extending the level of analytical control previously available only in two-dimensional effective models.

\end{abstract}
\maketitle

\noindent\textit{\textbf{Introduction}.}  
The formation of spacetime  singularities in gravitational collapse is a robust prediction of classical general relativity \cite{PhysRevLett.14.57, PhysRevLett.15.689, PhysRevLett.17.444, PhysRevLett.17.445, Hawking:1970zqf, Senovilla_2015}. Understanding their fate requires incorporating quantum effects \cite{cmp/1103899181, cmp/1103899393, PhysRevD.12.1519}, which are expected to become important in regions of large curvature. Even at the semiclassical level, where  matter fields are quantized while spacetime remains classical \cite{DEWITT1975295, birrell_davies_1982, Fulling_1989, Wald:1995yp, parker_toms_2009, doi:10.1142/S0217751X13300238}, the vacuum energy of quantum fields in Einstein's equations, $G_{ab}=8\pi G\langle 0|T_{ab}|0\rangle$, can  modify the  causal structure of spacetime. In particular,  violations of classical energy conditions are widely known in quantum field theory \cite{Casimir:1948dh, Epstein:1965zza, PhysRev.184.1272, Davies:1977yv, PhysRevD.51.4277, doi:10.1142/9789812704016_0056}  rendering the classical singularity theorems inapplicable.
While a complete understanding  ultimately requires a full theory of quantum gravity, the semiclassical framework provides a controlled setting in which quantum effects of matter fields can be consistently incorporated \cite{delRio2025_backreaction, Juarez-Aubry_2026}, and is therefore expected to capture essential features of black-hole evaporation and offer valuable insight into the associated information loss issue \cite{PhysRevD.14.2460, Unruh_2017, Ashtekar:2022oyq}.

Resolving the semiclassical field equations, however, remains a major challenge. The computation of $\langle 0|T_{ab}|0\rangle$ requires renormalization of ultraviolet divergences in general curved spacetimes, which in four spacetime dimensions is technically challenging   even for free fields. Even though the renormalization problem can be  formulated rigorously within the Hadamard prescription \cite{robert_m__wald_1977, PhysRevD.14.2490, PhysRevD.17.946, cmp/1103904566, FULLING1981243, PhysRevD.17.946}, the resulting expressions for $\langle 0|T_{ab}|0\rangle$ depend on state-dependent  contributions \cite{Hadamard1923, ADLER1977279, PhysRevD.17.1477, ADLER1978294, CASTAGNINO198485,10.1063/1.526008, PhysRevD.34.1776, PhysRevD.34.2286, 10.1143/PTP.77.671, 10.1143/PTP.81.891, PhysRevD.78.044025, PhysRevD.93.044063} that are typically inaccessible in practice, as they require solving  field equations analytically,   evaluating infinite mode sums, and performing coincidence limits along geodesics. Considerable effort has therefore been devoted to developing more efficient computational schemes, combining analytical and numerical techniques, with explicit results in  cosmology  \cite{PhysRevD.9.341, FULLING1974176, PhysRevD.10.3905, Bunch_1978, PhysRevD.18.1844, 300b6d7c-10d8-3655-99b9-25256f6839ee, Bunch_1980, PhysRevD.36.2963, PhysRevD.41.3101, PhysRevD.89.044030, PhysRevD.90.084017, PhysRevD.91.064031, PhysRevD.95.105003, PhysRevD.103.105002, PhysRevD.106.085003} and stationary black holes (BHs) \cite{PhysRevD.21.2185, ELSTER1983205, PhysRevLett.53.403, PhysRevD.30.2532, PhysRevLett.70.1739, PhysRevD.51.4337, PhysRevD.66.124017, PhysRevD.94.125024,  PhysRevD.96.105020, PhysRevD.106.065023, PhysRevD.91.104028, PhysRevD.94.044054,  PhysRevLett.117.231101, PhysRevD.95.025007, PhysRevLett.118.141102, PhysRevD.97.024033, PhysRevLett.124.171302,  PhysRevD.104.024066, PhysRevLett.129.261102, PhysRevD.111.085009, PhysRevD.108.125004, PhysRevD.109.105022}. Yet, these approaches  rely on mode-sum decompositions on fixed backgrounds and numerical regularization methods, which obscure the structure of $\langle T_{ab}\rangle$ and hinder the analysis of backreaction. As a result, fully self-consistent semiclassical solutions describing  collapsing stars remain largely out of reach.

In view of these difficulties, lower-dimensional effective models have often been employed to study spherically-symmetric gravitational collapse, justified within the $s$-wave or Polyakov approximations \cite{POLYAKOV1981207, PhysRevD.45.R1005, SUSSKIND1992123, PhysRevD.46.3444, PhysRevD.47.2446, MUKHANOV1994283, doi:10.1142/p378, PhysRevLett.106.161303, PhysRevD.83.044040}. The key simplification is that, in two spacetime dimensions, an exact expression for $\langle 0|T_{ab}|0\rangle$ can be derived from conformal symmetry \cite{ceb8dc1c-d11b-3e0b-9c15-54c356f2bc68, 0fb308e0-b7a9-3288-b0cc-1f935580a086, PhysRevD.13.2720, PhysRevD.16.1712, Bunch_1978, WALD1978472, doi:10.1142/p378}, providing valuable insight into black-hole evaporation and backreaction  \cite{PhysRevD.13.2720, PhysRevD.15.3054, PhysRevD.23.2813,  doi:10.1142/p378,  PhysRevD.73.104023,   Ju_rez_Aubry_2018, PhysRevLett.120.061102, PhysRevD.101.064059, Barceloetal, PhysRevD.106.124006, PhysRevD.111.045025}. These models successfully reproduce the $\ell=0$ contribution to the Hawking radiation \cite{MUKHANOV1994283, doi:10.1142/p378}. However, the extent to which such results capture the actual four-dimensional dynamics remains unclear, as quantization does not commute with dimensional reduction \cite{PhysRevD.61.024021, PhysRevD.62.044033}.

The framework developed here provides a higher-dimensional counterpart that retains analytical control while remaining applicable to genuine 4D spacetimes. Unlike conventional implementations of the Hadamard prescription, which rely on mode sums to reconstruct the state-dependent contributions, our reformulation makes the state-dependent data explicitly parametrizable and reduces the computation of $\langle 0|T_{ab}|0\rangle$ to the solution of a universal first-order covariant equation, with the choice of quantum state entering through a source term. This formulation allows explicit closed-form expressions in symmetric 3+1-dimensional curved spacetimes  and renders the semiclassical Einstein equations tractable. As an application, we construct explicit families of renormalized stress tensors in Schwarzschild spacetime, including Unruh-like and Boulware-like states, and obtain metric solutions describing evaporating BHs. In particular, the Unruh-like states reproduce the  Hawking flux at future null infinity together with its negative-energy ingoing counterpart at the horizon.

\noindent\textit{\textbf{Closed-form representation of the renormalized stress-energy tensor}.}
Let $\Phi$ be a free scalar field of mass $m$ and curvature coupling $\xi$ on a  globally hyperbolic $d$-dimensional  spacetime  with metric $g_{ab}$, covariant derivative $\nabla_a$,  Ricci tensor $R_{ab}$ and scalar curvature  $R=g^{ab}R_{ab}$. The stress-energy  operator is obtained by promoting the classical expression, 
\bea
T_{a b}&=&\nabla_{a} \Phi \nabla_{b} \Phi-\frac{1}{2} g_{a b}(\nabla \Phi)^{2}-\frac{1}{2} m^{2} g_{a b} \Phi^{2} \nonumber\\
&&+\xi\left(R_{a b}-\frac{1}{2}g_{ab}R-g_{a b} \nabla_c\nabla^c +\nabla_{a} \nabla_{b} \right)\Phi^{2} \label{Tab}\, ,
\eea
to the quantum theory. Due to the distributional nature of quantum fields  \cite{osti_4606723, 76b9be9a-dc2a-31f1-86b5-18215df947a7}, expectation values of  local products such as $\langle \hat\Phi(x)\hat \Phi(x)\rangle$ are ill-defined mathematically. Physically, this reflects the presence of ultraviolet divergences. 
A  renormalization scheme consistent with locality, covariance and causality is  provided by the Hadamard prescription \cite{robert_m__wald_1977}, in which these divergences are  regularized by covariant point-splitting  \cite{PhysRevD.14.2490, PhysRevD.17.946},   so that the renormalized  tensor, defined by
\bea
\left\langle T_{a b}\right\rangle_{\mathrm{}}:=\lim _{x^{\prime} \rightarrow x}D_{ab'}\left[\langle\{\hat{\Phi}(x) \hat{\Phi}(x')\}\rangle-\langle\{\hat{\Phi}(x) \hat{\Phi}(x') \} \rangle_{\mathrm{sing}}\right]\, , \nonumber
\eea 
 is well-defined. Here $D_{ab'}$ is a bi-differential operator constructed from (\ref{Tab}), and $\{\cdot,\cdot\}$ denotes symmetrization. The subtraction term removes the  short-distance singular structure, while the corresponding regularized divergences can be absorbed into the coupling constants of the semiclassical equations \cite{PhysRevD.17.946, PhysRevD.78.044025}.
 
  Although the full two-point function $\langle \hat\Phi(x)\hat \Phi(x')\rangle$ depends on the  quantum state, the short-distance singular part $\langle\{\hat{\Phi}(x) \hat{\Phi}(x') \} \rangle_{\mathrm{sing}}$ only depends on the local spacetime geometry for Hadamard states \cite{Hadamard1923}, making this prescription useful in any curved spacetime within this class. Explicit expressions for $\langle 0|T_{ab}|0\rangle$ can then be obtained  in a generic $d$-dimensional spacetime  \cite{ADLER1977279, PhysRevD.17.1477, ADLER1978294, CASTAGNINO198485,10.1063/1.526008, PhysRevD.34.1776, PhysRevD.34.2286, 10.1143/PTP.77.671, 10.1143/PTP.81.891, PhysRevD.78.044025, PhysRevD.93.044063},
\bea
\left\langle T_{a b}\right\rangle&=&\frac{\hbar \alpha_{d}}{2}\Big[\frac{(1-2 \xi)}{2} \nabla_{a} \nabla_{b} \omega-\omega_{a b}+\frac{\left(2 \xi-\frac{1}{2}\right)}{2} g_{a b} \square \omega\nonumber\\
 &&\hspace{1cm}+\xi R_{a b} \omega-\epsilon_d g_{a b} v_{1}\Big]+ \Theta_{ab}\label{T1}\, ,
\eea
where $\alpha_{d}=\frac{\delta_{d2}}{2\pi}+(1-\delta_{d2})\frac{\Gamma(d/2-1)}{(2\pi)^{d/2}}$, $\epsilon_d=\frac{1+(-1)^d}{2}$,   $v_1$ is a local geometric scalar encoding the trace anomaly, and $\Theta_{ab}$ encodes  renormalization ambiguities (e.g.  renormalization scale)   fixed by  local curvature terms \cite{PhysRevD.78.044025}. The dependence on the quantum state only appears in 
\bea
\omega(x)=\lim _{x^{\prime} \rightarrow x} W\left(x, x^{\prime}\right)\, ,\quad  \omega_{a b}=\lim _{x^{\prime} \rightarrow x} \nabla_a\nabla_bW\left(x, x^{\prime}\right)\nonumber\, ,
\eea
where $W\left(x, x^{\prime}\right)\propto \langle\{\hat{\Phi}(x) \hat{\Phi}(x')\}\rangle-\langle\{\hat{\Phi}(x) \hat{\Phi}(x') \} \rangle_{\mathrm{sing}}$. 

Determining these quantities requires reconstructing the  two-point function, typically  via mode-sum computations. Since this is inaccessible in most practical situations, we seek a reformulation that isolates the minimal state-dependent information required to determine $\langle T_{ab}\rangle$. In the standard  framework, this information is redundantly encoded in $\omega$, $\omega_{ab}$, which satisfy the constraints  
\bea
\omega^a_a&=&(m^2+\xi R)\omega -\epsilon_d(d+2) v_1\, ,\label{c11}\\
 \omega_{a \, \, \, \,; \rho}^{\, \, \, \rho} & = & \frac{1}{4}\nabla_a \square \omega+ \frac{1}{2} R_{a}^{\, \, \, c}\nabla_c \omega  +\frac{1}{2}\xi \omega\nabla_a R  -\epsilon_d \nabla_a v_1 \, . \label{c22}
\eea
These relations 
show that  not all components of $\omega_{ab}$ are independent. The key observation is that these constraints can be solved by decomposing $\omega_{ab}$ as
\bea
\omega_{a b}&=&\Big(\frac{1}{2}-\xi\Big)\Big(\nabla_{a} \nabla_{b} \omega-\frac{1}{d} g_{a b} \square \omega\Big)+\xi R_{a b} \omega\nonumber\\
&&+g_{ab}\Big(\frac{m^2}{d}\omega-\epsilon_d\frac{d+2}{d}  v_{1}\Big)+\omega_{ab}^T\, ,
\eea
where $\omega_{ab}^T$ is symmetric and traceless. 
In terms of this decomposition, the standard expression (\ref{T1}) for $\langle T_{ab}\rangle$ can be recast in a much simpler and more transparent form,
\bea
\langle 0|{T}_{a b}| 0\rangle=  \frac{\hbar \alpha_d}{d}\left[ g_{ab}\, v_1^{\omega} -\frac{d}{2}\omega^T_{ab}\right] + \Theta_{ab} \label{final3}\, ,
\eea
where $v_1^\omega=\epsilon_d v_1-\frac{m^2}{2}\omega+\frac{(d-1)}{2}(\xi-\xi_c(d))\Box \omega$, with $\xi_c(d)=\frac{d-2}{4(d-1)}$ the conformal coupling in $d$ dimensions. Furthermore, the constraints (\ref{c11}) and (\ref{c22}) reduce to
\bea
\omega^{Ta}_a=0\, ,\quad  \nabla^a\omega^T_{ab}=\frac{2}{d}\nabla_b v_1^\omega\, , \label{constraints}
\eea
so that the remaining freedom is encoded in the divergence-free sector $\omega_{ab}^{TT}$ of the traceless tensor $\omega^T_{ab}$.

The decomposition (\ref{final3}) makes the role of the quantum state in the renormalized stress-energy tensor completely explicit. The scalar function $\omega(x)$ is freely specifiable and acts as an effective source for  $\omega^T_{ab}$ through the first-order equation (\ref{constraints}). This observation naturally suggests taking $\omega(x)$, rather than the full biscalar $W(x,x')$, as the fundamental state-dependent variable. Determining $\langle T_{ab}\rangle$ then reduces to prescribing $\omega$ and solving the universal first-order covariant equation (\ref{constraints}), thereby replacing the reconstruction of the full two-point function using mode-sum computations by a local and geometrically transparent procedure.

For  conformal  fields in $1+1$ dimensions ($m=\xi=0$), this construction reproduces exactly the well-known expression derived from conformal symmetry \cite{ceb8dc1c-d11b-3e0b-9c15-54c356f2bc68, 0fb308e0-b7a9-3288-b0cc-1f935580a086, PhysRevD.13.2720, PhysRevD.16.1712, Bunch_1978, WALD1978472, doi:10.1142/p378}. In this case $v_1^\omega=\frac{R}{12}$ so the source term $\omega$ drops out, and Eq. (\ref{constraints}) can be integrated exactly with $\omega_{ab}^T= \omega_{ab}^{TT}-\frac{1}{3}[\nabla_{a}\nabla_b \rho +\nabla_a \rho \nabla_b\rho-\frac{1}{2}g_{ab}(\Box \rho+\nabla_c\rho \nabla^c\rho)]$, for any scalar  satisfying $\Box \rho=-\frac{R}{2}$. For $\omega_{ab}^{TT}=0$, this yields the standard result, without invoking conformal methods. 

Notably, this approach does not rely on the  special conformal symmetry  of two-dimensional spacetimes in order to get an explicit result, and therefore extends naturally to higher dimensions. In general $3+1$-dimensional backgrounds, a closed local expression for $\omega^T_{ab}$ is not available. However, with the ansatz $\omega_{ab}^T=\omega_{ab}^{TT}+\nabla_{(a} F_{b)}-\frac{1}{4}g_{ab}\nabla_c F^c$, Eq. (\ref{constraints}) produces $(g^b_a\Box+\frac{1}{2}\nabla_a \nabla^b+R_a^b)F_b=\nabla_b v_1^\omega$. This is a normally hyperbolic differential equation, and therefore  admits a formal solution in terms of Green functions $G_{ab}(x,x')$ \cite{doi:10.1142/S0217751X13300238}.
\bea
F_{b}(x)=\int d^4x'\sqrt{-g(x')}\, G_{b}^{\, \, \, a^{\prime}}\left(x, x^{\prime}\right) \nabla_{a^{\prime}} v_1^{\omega}\left(x^{\prime}\right) \, .
\eea
 In symmetric spacetimes, Eq. (\ref{constraints}) can be integrated explicitly, leading to closed-form expressions for $\langle T_{ab}\rangle$. We illustrate this in the next section. 
 
\noindent\textit{\textbf{Explicit formulas for $\langle 0| T_{ab}|0\rangle$ in a Schwarzschild BH}.}  
We consider a 3+1-dimensional spherically symmetric spacetime. To solve Eq. (\ref{constraints}), we introduce a congruence  of  observers  with unit timelike 4-velocity $U^a$ and vanishing twist. This provides a foliation by spacelike hypersurfaces orthogonal to $U^a$, with induced spatial metric $h_{ab}=g_{ab}+U_aU_b$. The kinematics of the congruence is characterized by the 3+1 decomposition $\nabla_a U_b = - U_a a_b + \frac{1}{3}\theta h_{ab} + \sigma_{ab}$, where $a_a=U^b\nabla_b U_a$ is the observers' acceleration, $\theta=\nabla_a U^a$ their expansion, and $\sigma_{ab} = h^n_a h^m_b \nabla_{(m}U_{n)}-\frac{\theta}{3}h_{ab}$ the shear. Spherical symmetry allows the further decomposition $h_{ab}=X_a X_b + q_{ab}$, with $q_{ab}$  the metric on the 2-spheres and $X^a$  the normal unit radial vector, in such a way that the acceleration is purely radial, $a_a=A X_a$, and $\sigma_{cd}X^d\, q^c_{a}=0$.

 We parametrize the traceless, symmetric tensor $\omega_{ab}^T$  as
\bea
 \omega^T_{ab} =2 \alpha_1 U_{(a}X_{b)} + \alpha_2 U_a U_b + (\alpha_2-2\alpha_3) X_a X_b + \alpha_3 q_{ab}  \nonumber
\eea
which is the most general form compatible with spherical symmetry. Substituting into Eq.~(\ref{constraints}) yields two independent equations $E_1=E_2=0$, where
\bea
 E_1&=&   -2(A+\beta)\alpha_1 -\left(\frac{4\theta}{3} +\sigma_{ab}X^aX^b\right)\alpha_2 \\
&&+3 \alpha_3 \sigma_{ab}X^aX^b -X^a\nabla_a \alpha_1 - U^a\nabla_a \alpha_2- \frac{1}{2}U^a\nabla_a v_1^{\omega}\, ,\nonumber\\ 
E_2 &=& 2(A+\beta)\alpha_2-(2A+6\beta)\alpha_3+ \left(\frac{4\theta}{3}+\sigma_{ab}X^aX^b\right)\alpha_1\nonumber\\
&&+U^a\nabla_a\alpha_1+X^a\nabla_a(\alpha_2-2\alpha_3)- \frac{1}{2} X^a\nabla_a v_1^{\omega}\, ,
\eea
with $\beta=\frac{1}{2}q^{ab}\nabla_a X_b$. These equations must be supplemented with boundary conditions.  Physically, different choices  correspond to different realizations of the quantum state. We  illustrate this idea  in Schwarzschild spacetime for conformal fields ($m=0$, $\xi=1/6$).

\textit{Unruh-like states.} For BHs  the Unruh vacuum plays a prominent role \cite{PhysRevD.14.870}. This state is stationary, regular on the future horizon and effectively reproduces Hawking radiation at future null infinity. Regularity motivates the use of a  Painlev\'e-Gullstrand frame, physically associated to a congruence of infalling observers,
\bea\label{freefallingobs}
    U= \frac{1}{1-2M/r}\partial_t  -\sqrt{\frac{2M}{r}} \, \partial_r\, , \quad X=-\frac{\sqrt{\frac{2M}{r}}  }{1-2M/r}\partial_t +\partial_r \  , \nonumber
\eea
while stationarity requires only radial dependence for the coefficients $\alpha_i$.
In this basis, the combination $E_1+\sqrt{\frac{2M}{r}}E_2=0$  reduces to
\bea
(\alpha_3-\alpha_2)'+\frac{3(\alpha_3-\alpha_2)}{2r}=-\frac{2(M+r)\alpha_1+(2M+r)r\alpha_1'}{2\sqrt{2Mr^3}}\nonumber\, ,
\eea
which can be integrated  as $\alpha_3-\alpha_2=\frac{C}{r^{3/2}}-\frac{\alpha_1 (2M+r)}{2\sqrt{2Mr}}$. The integration constant $C$ is then fixed by imposing the asymptotic Hawking flux at future null infinity $\langle T_{uu}\rangle\sim \frac{L}{4\pi r^2}+O(r^{-4})$, $\langle T_{vv}\rangle \sim O(r^{-4})$, yielding $C=-\frac{\pi L}{\sqrt{2M}}$ along with the fall-off condition $\alpha_1 \sim -\frac{2\pi L}{r^2}
\sum_{n=0}^{4} (n+1) ({\frac{2M}{r}})^{n/2}+O(r^{-9/2})$. The remaining combination   $E_1-\sqrt{\frac{2M}{r}}E_2=0$ then reads
\bea
\alpha_3'+\frac{4\alpha_3}{r}= \frac{M^2}{5\, r^7}+\frac{\pi L}{ 2\sqrt{2 M\, r^5}}+\frac{(5r-2M)\,\alpha_1}{4 \sqrt{2M\, r^3}}+\frac{(r-2M)\,\alpha_1'}{2\sqrt{2 M\, r}}\, , \nonumber
\eea
and can also be integrated exactly, yielding eventually
\bea
\alpha_3(r) =& -\frac{M^2}{10\, r^6} + \frac{D}{2\, r^4}- \frac{1}{\sqrt{2 M} r^4} \int r^{5/2}\,(r-3M)\,\alpha_1(r)\,dr \nonumber\\
& + \frac{\pi L}{5 \sqrt{2M r^3}} + \frac{\sqrt{r}\,\alpha_1(r)}{2 \sqrt{2M}}\Big(1-\frac{2M}{r}\Big) \,,\\
\alpha_2(r) =& -\frac{M^2}{10\, r^6} + \frac{D}{2\, r^4}- \frac{1}{\sqrt{2M}r^4} \int r^{5/2}\,(r-3M)\,\alpha_1(r)\,dr \nonumber\\
& + \frac{6\pi L}{5 \sqrt{2M r^3}}+ \frac{\sqrt{r}\,\alpha_1(r)}{\sqrt{2M}} \,.
\eea
Here $\alpha_1(r)$ remains a free function encoding residual  dependence on the quantum state, and $D$ is an integration constant. Remarkably, this entire family of solutions  satisfies $\langle T_{uu}\rangle |_{r\to 2M}= 0$, $\langle T_{vv}\rangle |_{r\to 2M}= -\frac{L}{4\pi (2M)^2}$ at the horizon for all $\alpha_1$, reproducing the expected ingoing negative-energy  flux responsible for the BH  {\it thermal} evaporation \cite{PhysRevLett.46.382, 10.1143/PTP.63.1217, HAJICEK19809, Balbinot_1984, PhysRevLett.73.2805, PhysRevD.52.5857, PCWDavies_1978}. While this behavior is well known analytically in 1+1 dimensions \cite{PhysRevD.13.2720, PhysRevD.15.2088}, in 3+1 dimensions it has so far been supported primarily by numerical calculations \cite{ELSTER1983205, PhysRevLett.117.231101, PhysRevD.95.025007} and indirect arguments \cite{PhysRevD.21.2185}. 

\textit{Boulware-like states.} For the exterior vacuum of static stars,  the Boulware vacuum is the physically relevant  state \cite{PhysRevD.11.1404}. It  is characterized by reproducing the asymptotic Minkowski behaviour near spatial infinity.  The natural congruence to choose is that of static observers, 
\bea
    U = \frac{1}{\sqrt{1-2M/r}} \, \partial_t \, , \qquad X=\sqrt{1-2M/r}\, \partial_r \  .
\eea
In this case $E_1=0$ is a decoupled equation for $\alpha_1$, 
\bea
2(r-M)\alpha_1+r(r-2M)\alpha'_1=0\, ,
\eea
which integrates to $\alpha_1=\frac{C}{r^2(1-2M/r)}$. Asymptotic flatness at spatial infinity requires  $\langle T_{ab}\rangle\to 0$ faster than $r^{-2}$, which enforces $C=0$. Thus $\alpha_1=0$, and $\langle T_{ab}\rangle$  contain no flux components for static observers, as expected for the Boulware state. Defining now the auxiliary  variable, $y=3\alpha_3-\alpha_2$,  the remaining equation $E_2=0$,
\bea
\big(\alpha_2' - 2\alpha_3'\big)+ \frac{2(r-M)}{r(r-2M)}\alpha_2+ \frac{2(5M - 3r)}{r(r-2M)}\alpha_3=-\frac{M^2}{5 r^7} \,,\nonumber
\eea
 can be integrated for both $\alpha_2$ and $\alpha_3$ as
\bea
\alpha_3 = \frac{I(r)}{(1-2M/r)^2}+ y(r)\, , \quad \alpha_2 =\frac{3I(r)}{(1-2M/r)^2}+ 2y(r)\, , \nonumber
\eea
with
$I(r)=\frac{M^2(21M^2-24M r+7r^2)}{210r^8} +  \int \frac{2(r-3M)(r-2M) y(r)dr}{r^3}$.
For a broad class of  $y(r)$, the solution reproduces the standard Boulware divergence with thermal structure $\langle T_a^b\rangle \sim (1-2M/r)^{-2}\mathrm{diag}(-3,1,1,1)$  near the horizon \cite{PhysRevD.21.2185, PhysRevLett.94.061301, PhysRevD.107.085023}. Interestingly, suitable choices of $y(r)$ can yield configurations that remain regular at the horizon, suggesting a broader class of static semiclassical solutions beyond the standard Boulware vacuum.

\noindent\textit{\textbf{Analytical solution of the semiclassical Einstein's equations}.}  
The explicit families of renormalized stress tensors obtained in the previous section render the semiclassical Einstein equations analytically tractable. In particular, the Unruh-like family provides an effective description of Hawking evaporation while avoiding both the detailed modelling of the collapsing geometry and the mode-sum constructions usually required in four-dimensional QFT in curved spacetime. Here we use this framework to construct an explicit quasi-stationary solution describing the backreaction of Hawking radiation on an evaporating Schwarzschild BH, valid throughout the semiclassical regime outside the region where curvature becomes Planckian. For definiteness, we consider the minimal asymptotic realization of the Unruh-like family compatible with the Hawking flux, $\alpha_1 = -\frac{2\pi L}{r^2}\sum_{n=0}^{4} (n+1) ({\frac{2M}{r}})^{n/2}$, which reproduces the required asymptotic behaviour at future null infinity.

Previous works have established the consistency of the quasi-stationary picture  using energy-conservation arguments, effective near-horizon flux models, Vaidya-type geometries, or models inspired by two-dimensional conformal results \cite{HAJICEK19809,PhysRevLett.46.382,Balbinot_1984,PhysRevLett.73.2805,PhysRevD.52.5857}. The construction below sharpens this picture: 
the renormalized stress tensor is given explicitly in four dimensions in a Unruh-like vacuum, and  the semiclassical  equations are solved directly, yielding the backreacted geometry in closed form.

We start from the Bardeen ansatz \cite{PhysRevLett.46.382}
\bea \label{bardeen}
    ds^2 = -\left[ 1- \frac{2 m(v,r)}{r}\right]e^{2\Psi(v,r)}dv^2 + 2 e^{\Psi(v,r)}dv \,dr + r^2 d\Omega^2 \ ,\nonumber
\eea
where $m(v,r)$ is the Misner-Sharp (or Hawking) quasi-local mass function \cite{PhysRev.136.B571, 10.1063/1.1664615}. In these coordinates, the semiclassical field equations reduce to
\bea
\partial_r m=-4\pi r^2 \langle T_v^v\rangle\, ,\quad  \partial_v m=4\pi r^2 \langle T_v^r\rangle\, , \quad\partial_r \Psi=4\pi r e^{\Psi} \langle T_r^v\rangle \, . \nonumber
\eea
Substituting the stress tensor obtained in the previous section and working consistently to leading order in $\hbar$, these equations can be integrated {\it exactly}, giving 
\begin{widetext}
\bea
\Psi&=&\hbar\tilde\Psi(v)+\frac{2}{3}\hbar L \left[3\log\!\Big(r (\sqrt r+\sqrt{2M})^2\Big)-\frac{2\sqrt{2M}}{r^{3/2}(\sqrt{r}+\sqrt{2M})}\left(5\sqrt{2}\,M^{3/2}+11M\sqrt{r}+\frac{15}{2}\sqrt{2M}\,r+3r^{3/2}\right) \right]\, ,\nonumber\\
m &=&M-\hbar L v + \frac{\hbar  \left[45D r^2-2M^2+360\pi L r^2\left(r^2-4M^2-12M^2\log r+Mr\log\! \, r(\sqrt r+\sqrt{2M})^2+\frac{80\sqrt2\,M^{5/2}}{3\sqrt r}-2M \sqrt{2Mr}\right) \right]}{180\pi r^3}\, ,\nonumber
\eea
\end{widetext}
where $\tilde\Psi(v)$ is an arbitrary function that can be absorbed into a redefinition of the advanced time coordinate $v$.   The  solution satisfies $G_{ab} [g] = 8\pi \langle T_{ab}\rangle_{\rm schw} + {O}(\hbar^2)$ without any further approximation at this order, and remains regular everywhere except at  $r=0$.  

Crucially, the quantum corrections  modify the causal character of the apparent horizon. In these coordinates the apparent horizon is defined by the spheres  $\{v_{ah},r_{ah}\}$ satisfying $2m(v_{ah},r_{ah})=r_{ah}$. The mass of these spheres decreases as $m(v_{ah},r_{ah})=M-\hbar L (v_{ah}-v_{ah}^0)+O(\hbar^2)$, as expected from thermal evaporation. The  normal vector to the apparent horizon has norm $g^{ab}\nabla_a(r-2m)\nabla_b(r-2m)=4L \hbar+O(\hbar^2)>0$, showing that the apparent horizon becomes timelike, rather than null, during the evaporation process. Consequently, information crossing the apparent horizon is not permanently hidden behind a null causal boundary \cite{PhysRevD.14.2460}. Although a complete assessment of information recovery requires going beyond the present semiclassical treatment, the backreacted geometry avoids the standard picture in which information remains forever trapped behind an event horizon.

Introducing, at leading order in $\hbar$, the retarded time coordinate $u=v-2r_*+O(\hbar)=v-2r-4M \log(r/2M-1)+O(\hbar)$, the Bondi mass at future null infinity yields $M_B(u):=\lim_{r\to \infty}m(v(u,r),r)=M-\hbar L (u-u^0)+O(\hbar^2)$, which decreases linearly with retarded time, precisely as expected from the Hawking luminosity. This provides an explicit four-dimensional realization of semiclassical black-hole evaporation obtained directly from the semiclassical Einstein equations.


\noindent\textit{\textbf{Conclusions}.} 
In this work we have reformulated the  Hadamard prescription in a form that yields closed-form expressions for the renormalized stress-energy tensor in arbitrary $3+1$-dimensional curved spacetimes. The central result is Eq. (\ref{final3}), which reduces the determination of $\langle T_{ab}\rangle$ to specifying the state-dependent  function $\omega$ and solving a universal first-order covariant equation.

This reformulation drastically simplifies the standard framework by isolating the minimal state-dependent information required to construct $\langle T_{ab}\rangle$, replacing the usual mode-sum construction by a local and geometrically transparent procedure. In 1+1 dimensions it allows us to recover the well-known Polyakov expression  without invoking  conformal symmetry. As a first non-trivial application in 3+1 dimensions, we have constructed  families of exact solutions for the renormalized stress tensor  of conformal fields in Schwarzschild spacetime, encompassing and generalizing the Unruh and Boulware states. In particular, the Unruh-like family reproduces the Hawking flux at future null infinity together with its negative-energy ingoing counterpart at the BH horizon, independently of the residual functional freedom in the quantum state. This, in turn, allowed us to obtain an explicit quasi-stationary solution of the semiclassical Einstein equations describing black-hole evaporation in 3+1 dimensions. These results can be straightforwardly extended to include mass and other  curvature couplings. 

Beyond Schwarzschild spacetime, our formalism applies naturally to a wide range of geometries. In particular, in Friedmann-Lemaitre-Robertson-Walker spacetimes it recovers the standard expression for $\langle T_{ab}\rangle$ in the conformal vacuum \cite{Parker1979, 10.1143/PTP.81.891} by adopting a congruence compatible with homogeneity and isotropy. Likewise, the same construction extends directly to Reissner-Nordstrom spacetimes, providing a natural framework to study the semiclassical instability of Cauchy horizons within a fully four-dimensional framework.

More generally, our framework renders the semiclassical Einstein equations formally tractable. Once prescriptions for the freely specifiable state-dependent data $(\omega, \omega_{ab}^{TT})$ are provided, the system can be formulated as a Cauchy initial value problem. For instance, choosing $(\Box-m^2-\xi R)\omega=\gamma v_1$, with $\gamma\in\mathbb{R}$, together with the minimal condition $\omega_{ab}^{TT}=0$, leads to a closed system of evolution equations for the variables $(g_{ab},F_a,\omega)$. With suitable initial data on a spacelike Cauchy hypersurface $\Sigma$, and treating higher-order derivative contributions through the method of perturbative constraints \cite{PhysRevD.41.3720, PhysRevD.43.3308, PhysRevD.47.1339}, the problem is well posed and amenable to numerical implementation using standard techniques in numerical relativity \cite{Alcubierre:1138167}. In this perspective, the special class of Hadamard states recently proposed in Ref.~\cite{JuarezAubry2026} is naturally recovered in this framework by restricting to  $\omega_{ab}^T=0$. More broadly, the present framework extends to realistic four-dimensional spacetimes the level of analytical control previously available only in two-dimensional effective models, opening new avenues for the study of quantum backreaction.

{\bf \em Acknowledgments.} 
We are grateful to Ivan Agullo, Julio Arrechea, Pablo Blasco-Gil, Alessandro Fabbri, Jose Navarro-Salas, Gonzalo Olmo and Nicolas Sanchis-Gual for valuable discussions and comments. We also thank the participants of the \href{https://indico.global/event/16783/}{EREP 2026} conference for helpful feedback on an earlier presentation of this work.
ADR acknowledges support through {\it Atraccion de Talento Cesar Nombela} grant No 2023-T1/TEC-29023, funded by Comunidad de Madrid (Spain). 
FJM-G is supported by the Ministerio de Ciencia, Innovaci\'on y Universidades, Ph.D. fellowship, Grant No. FPU22/02528.
This work has also been supported by the Spanish Grant PID2023-149560NB- C21 funded by MCIU /AEI/10.13039/501100011033 / FEDER, UE and by the Grant CEX2023-001292-S funded by MCIU/AEI.  The paper is based upon work from COST Action CaLISTA CA21109 supported by COST (European Cooperation in Science and Technology).

\appendix

\bibliography{references}

\end{document}